\documentclass[a4paper,11pt]{article}
\usepackage{aaskaiid}

\title{Polarimetry of Ultraluminous X-ray sources}
\ShortTitle{Polarimetry of Ultraluminous X-ray sources}

\author[1, 2]{Haruka Sakemi}
\ShortName{Sakemi, H.} 

\affiliation[1]{Graduate School of Sciences and 
Technology for Innovation, Yamaguchi University, 1677-1 Yoshida, Yamaguchi, 753-0841, Japan}
\emailAdd{sakemi@yamaguchi-u.ac.jp}
\affiliation[2]{Nobeyama Radio Observatory, National Astronomical Observatory of Japan (NAOJ), National Institutes of Natural Sciences
(NINS), 462-2, Nobeyama, Minamimaki, Minamisaku, Nagano 384-1305, Japan}

\abstract{Ultraluminous X-ray sources (ULXs) are non-nuclear X-ray emitters exceeding the Eddington luminosity, offering key insights into extreme accretion physics. Their origin is attributed to either sub-Eddington accretion onto intermediate-mass black holes (IMBHs) or supercritical accretion onto stellar-mass compact objects in binary systems. A subset of ULXs shows radio jets and surrounding bubbles, indicative of strong mechanical feedback, but no polarization detection has been reported to date because of limited sensitivity and angular resolution at multi-megaparsec distances.

Next-generation facilities such as the Square Kilometre Array (SKA) will enable polarimetric studies of ULX outflows and bubbles, providing vital information on their magnetic field structure and jet–ISM interactions. Using Holmberg II X-1 (3.39 Mpc) as a representative system, we evaluated the detectability of its polarized jet and bubble with the SKA-Mid AA4 configuration. Assuming a fractional polarization of 20\% and Band 2 observations centered at 1310 MHz, the expected polarized intensities are 260 $\mu$Jy beam$^{-1}$ for the jet and 20 $\mu$Jy beam$^{-1}$ for the bubble. A Holmberg II X-1–like jet is detectable within 10 hours even at $\sim$ 10 Mpc, while bubble polarization requires $\sim$ 100 hours. Although resolving fine structures is feasible only for nearby ULXs, future SKA-VLBI capabilities will help overcome this limitation and enable detailed mapping of magnetic field morphology.

These results indicate that the SKA will enable the first detection of polarized emission from ULXs, advancing our understanding of magnetic field geometry, jet formation, and feedback in super- and sub-Eddington accretors.

}

\begin{document}
\maketitle

\section{Ultraluminous X-ray Sources (ULXs)}
Ultraluminous X-ray Sources (ULXs) are non-active galactic nuclei (non-AGN) objects that exhibit extremely high X-ray luminosities exceeding the Eddington limit. Although the Eddington luminosity depends on the mass of the accretor, a commonly adopted criterion for identifying a ULX is whether its X-ray luminosity surpasses approximately $10^{39}$ erg s$^{-1}$, corresponding to the Eddington luminosity of a 10 $M_\odot$ compact object. To date, more than 1,800 ULX candidates have been reported \citep{walton2022}, the majority of which are located in nearby galaxies, particularly in star-forming environments.

Several scenarios have been proposed to explain the physical origin of ULXs. One possibility involves sub-Eddington accretion onto an intermediate-mass black hole (IMBH) with a mass of $10^{2}$–$10^{4}\,M_\odot$ \citep{taniguchi2000}. In such systems, the large accretor mass naturally enables X-ray luminosities above $10^{39}$ erg s$^{-1}$, in some cases reaching $10^{40}$–$10^{42}$ erg s$^{-1}$. Sources of this class are sometimes designated as Hyperluminous X-ray Sources (HLXs) and are regarded as strong IMBH candidates.

Alternatively, ULXs may consist of stellar-mass compact objects in binary systems with ordinary stars. In typical accretion states, the observed luminosity cannot easily exceed the Eddington limit; however, supercritical accretion flows provide a plausible mechanism. When the mass inflow rate at the outer accretion disk substantially exceeds the Eddington accretion rate, the innermost region develops into a geometrically and optically thick slim disk. Powerful, angular-momentum-carrying winds are then launched from the disk surface. When such a system is viewed nearly face-on or along the outflow axis, the observed X-ray emission can appear significantly super-Eddington \citep{fabrika2021}. Distinct spectral states observed among ULXs are often interpreted as being primarily dependent on viewing angle. High-mass X-ray binaries (HMXBs) are capable of sustaining the necessary high accretion rates due to intense stellar winds from their massive companions. Nevertheless, even low-mass X-ray binaries (LMXBs) may exhibit transient phases of enhanced accretion during which their luminosities temporarily exceed the Eddington limit, mimicking ULX behavior.

\citet{Bachetti2014} detected coherent pulsations with a period of 1.37 s from the ULX M82 X-2, demonstrating that neutron stars can serve as accretors in ULXs. Subsequent discoveries of additional pulsating ULXs (see Table 2 of \citealt{King2023}) have confirmed this population. The mechanisms by which neutron star binaries achieve super-Eddington luminosities have been widely discussed. 
Magnetohydrodynamic (MHD) simulations, including radiation-MHD models (e.g., \citealt{kawashima2016, kawashima2020}), indicate that radiation can escape laterally through the sides of accretion columns formed above the magnetized neutron star surface.
When the magnetic and rotational axes are misaligned, observers may directly view the column sides, where locally super-Eddington fluxes are produced.

In recent years, several ULX candidates have also been identified within the Milky Way. Many of these are microquasars, providing empirical support for the hypothesis proposed by \citet{Freeland2006}, who suggested that some ULXs represent microquasars viewed face-on (microblazars). The Galactic microquasar SS 433, one of the most active known systems, is likely an edge-on analog of a supercritically accreting ULX. The X-ray flux from SS 433 is heavily absorbed—presumably by disk winds—such that the source remains sub-Eddington in X-rays but exceeds the Eddington luminosity in the ultraviolet band. The HMXB microquasar Cygnus X-3 is likewise considered a ULX candidate \citep{Yang2023}. Moreover, LMXBs such as GRS 1915+105 and V4641 Sgr—which occasionally reach super-Eddington luminosities and, in the latter case, have been detected in gamma rays—are also regarded as potential Galactic ULXs.


Despite significant progress, the underlying nature of ULXs remains elusive. In particular, key open questions concern the role of magnetic fields in shaping outflows, launching jets, and mediating interactions with the surrounding interstellar medium. Magnetic fields are also expected to play a central role in particle acceleration processes, potentially allowing ULXs to act as sources of high-energy cosmic rays.

Radio polarization observations provide a unique probe of these phenomena. The detection of polarized synchrotron emission would directly reveal the degree of magnetic field ordering and its geometry, offering insight into jet collimation, turbulence, and shock structures. In particular, enhanced polarization is expected in shock-compressed regions, which are also candidate sites of efficient particle acceleration. Therefore, polarization measurements can provide direct constraints on the physical conditions and mechanisms responsible for cosmic-ray acceleration in ULX systems.

Furthermore, polarization observations across a wide frequency range enable the study of Faraday rotation and depolarization, allowing us to probe the magneto-ionic environment both within the ULX outflows and along the line of sight. Such measurements are essential for understanding the coupling between accretion processes, outflows, and their surrounding medium.

In this context, the sensitivity and frequency coverage of SKA will provide a transformative opportunity to detect and characterize polarized emission from ULXs for the first time, opening a new window on their magnetic and high-energy processes.

\section{Radio Observations of ULXs}
Radio emission has been reported from approximately fifteen ULXs. Twelve of these objects are located in galaxies within 15 Mpc, while the remaining three lie at distances greater than 33 Mpc, with the most distant source residing in a galaxy 98.4 Mpc away \citep{soria2010a, Webb2012, Mezcua2013c, Mezcua2015, Mezcua2018, Cseh2015b}. The fact that these systems are detected at such large distances implies that they are likely ULXs hosting intermediate-mass black holes (IMBHs). Powerful radio jets have been observed in several of these sources.

In relatively nearby ULXs (within $\sim$10 Mpc), radio observations have revealed bubble or nebular structures surrounding the source \citep{Cseh2012, Berghea2020, Urquhart2018, Soria2021, Gong2023, Beuchert2024}. Roughly half of these exhibit sizes comparable to, or slightly larger than, typical supernova remnants (SNRs), while the others extend beyond 100 pc in diameter. Even among those with SNR-sized bubbles, the radio surface brightness is one to two orders of magnitude higher than that of canonical SNRs. Two principal mechanisms have been proposed to account for the formation of ULX bubbles: photoionization driven by radiation from the central source, and shock ionization produced where ejecta from the central engine interact with the surrounding interstellar medium. In some well-studied systems, such as Holmberg IX X-1, observational evidence suggests that both processes operate simultaneously \citep{Berghea2020}.

Given that ULXs are sources of powerful outflows, considerable attention has been devoted to understanding the outflow launching mechanism and its connection to the accretion flow. Upcoming coordinated monitoring campaigns combining the Square Kilometre Array (SKA) with sensitive X-ray observatories are expected to shed new light on this relationship. Furthermore, the detection of ULX bubbles carries an additional significance—it offers a means to identify potential ULX candidates that are not recognized as such in X-rays. As discussed previously, typical ULXs are thought to be viewed nearly along the polar axis. In contrast, systems viewed edge-on may obscure the central regions with the accretion disk or winds, thereby preventing their classification as ULXs even when undergoing supercritical accretion. Nevertheless, since such systems can still inflate bubbles through mechanical feedback, sensitive future radio surveys have great potential to uncover hidden or misoriented ULX systems.

\section{Polarimetric Observations of ULXs}
Although radio emission has been reported from several ULXs, no significant detection of radio polarization has yet been reported. We note that polarization has been detected in the X-ray band for at least one ULX (e.g., \citealt{Majumder2024}), highlighting the potential importance of multi-wavelength polarimetric studies.
This non-detection in radio is likely due to the fact that the radio-emitting ULXs are located in galaxies at distances of several Mpc or more, where current instruments face sensitivity limitations. In addition, beam depolarization caused by insufficient angular resolution may further complicate such observations. Nevertheless, polarimetric studies of ULXs are of considerable interest, and forthcoming next-generation radio interferometers such as the Square Kilometre Array (SKA), including very long baseline interferometry (VLBI) observations, are expected to provide new opportunities in this domain.

One of the most promising targets is the detection of polarization from ULX outflows. Investigating whether the magnetic field vectors are aligned parallel or perpendicular to the outflow axis can reveal critical information on the flow geometry, turbulence, and magnetic field configuration. If polarization can be measured across a wide frequency range, the degree of depolarization at lower frequencies can further constrain turbulence and the magnetic field strength or orientation in intervening Faraday screens along the line of sight. Such measurements would offer valuable insight into the magnetic field structure at the base of jets and outflows. Additionally, monitoring the temporal variability of the polarized emission will be highly informative. Coordinated X-ray and radio polarimetric monitoring can clarify how polarization properties, including depolarization, evolve with accretion states, thereby elucidating the role of magnetic fields in outflow launching.

Another promising avenue is the polarization study of ULX bubbles or nebulae. As noted above, ULXs share many similarities with microquasars or may belong to the same class of objects. Microquasars have recently attracted attention as potential “PeVatrons,” capable of accelerating cosmic-ray particles to energies exceeding 10$^{15}$eV \citep{LHAASO2024}. Accordingly, ULXs may also serve as sources of high-energy cosmic rays. Possible acceleration sites include the regions where outflows interact with the surrounding interstellar medium (ISM), the outflow launching zone, and internal shocks within the outflow itself. In the case of shock-ionized ULX bubbles, magnetic field structures associated with shocks at the outflow–ISM interaction front may produce detectable polarized emission. Coordinated observations of the ISM environment around ULX bubbles, combined with gamma-ray measurements, will be essential for identifying such regions and constraining their particle acceleration processes.

Faraday tomography has recently emerged as a powerful tool in radio polarimetry, enabling the disentanglement of multiple Faraday-rotating components along the line of sight \citep{Burn1966, Brentjens2005}. By applying this technique, it should become possible to reconstruct the three-dimensional magnetic field structures associated with ULX outflows and bubbles, as well as those in the surrounding Faraday screens. Analyses of Faraday spectra, including their line widths and substructures, can probe the local turbulence around ULXs, offering a unique opportunity to investigate turbulence driven by localized high-energy phenomena in galactic environments.

In addition to beam depolarization, internal Faraday rotation within the emitting region may significantly affect the detectability of polarized emission from ULXs. The rotation measure (RM) can be estimated as RM $\sim$ 0.81 $n_e$ $B_{||}$ $L$, where $n_e$ is the thermal electron density in cm$^{-3}$, $B_{||}$ is the magnetic field component along the line of sight in $\mu$G, and $L$ is the path length in pc.

For ULX jets and surrounding bubbles, plausible physical conditions are $n_e$ $\sim$ 0.1--10 cm$^{-3}$, $B$ $\sim$ 10--100 $\mu$G, and $L$ $\sim$ 1--100 pc, depending on the region and environment (e.g., studies of microquasar jets and supernova remnants; see, e.g., \citealt{Bordas2009}; \citealt{Reynolds2012}). These values imply RM $\sim$ 10$^2$--10$^4$ rad m$^{-2}$. At the observing frequency of 1.3 GHz ($\lambda$ $\sim$ 0.23 m), this corresponds to a polarization angle rotation of $\delta\chi$ $\sim$ RM $\lambda^{2}$ $\sim$ 5--500 rad, indicating that strong internal Faraday rotation may occur in some cases.

Such large rotation measures can lead to significant internal depolarization, particularly in extended and turbulent regions such as ULX bubbles, where fluctuations in magnetic field and electron density along the line of sight are expected. However, polarization may still be detectable in localized regions where the magnetic field is more ordered. In particular, shock-compressed regions at the interface between the outflow and the surrounding ISM are expected to enhance the degree of magnetic field ordering, potentially producing detectable polarized emission even in otherwise depolarized environments. 

In addition, more compact and possibly less turbulent jet components may retain a higher degree of polarization overall. These effects introduce an observational bias, making polarization detection more favorable for compact jet features, while in bubbles the polarized emission may be confined to specific regions associated with shocks or coherent structures.

Observations at higher frequencies or with sufficient spectral resolution to perform Faraday tomography will be essential to mitigate depolarization effects and to disentangle complex Faraday structures along the line of sight.

\section{Prospects with AA4}
\begin{figure}[t]
    \centering
	\includegraphics[width=0.7\columnwidth]{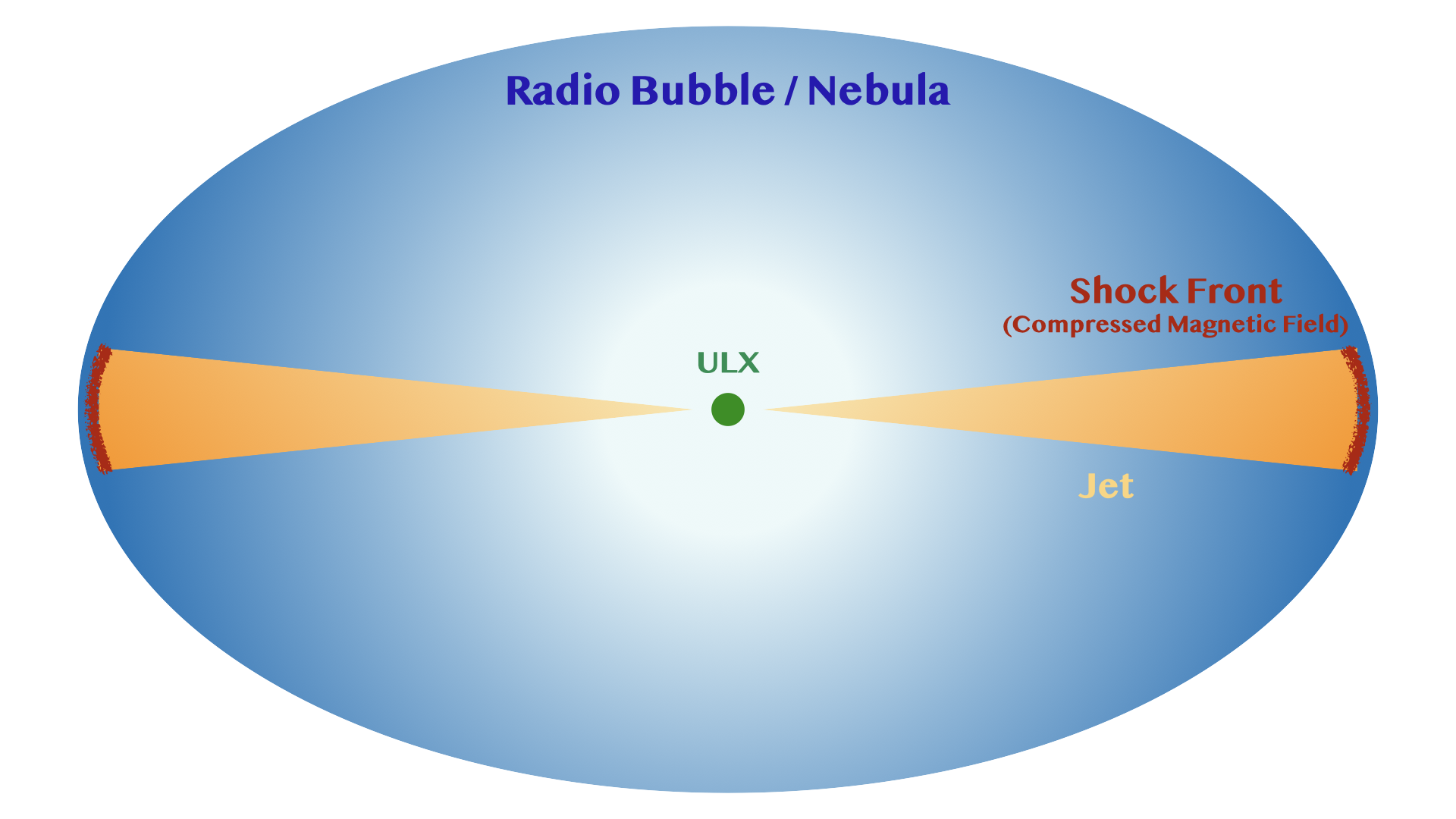}
    \caption{Schematic illustration of a ULX system showing the central source, bipolar jets, the surrounding radio bubble or nebula, and shock fronts at the interface with the interstellar medium. These regions represent potential sites of polarized emission and particle acceleration associated with ULX outflows.}
    \label{fig:schematic}
\end{figure}

\begin{figure}[t]
    \centering
	\includegraphics[width=0.7\columnwidth]{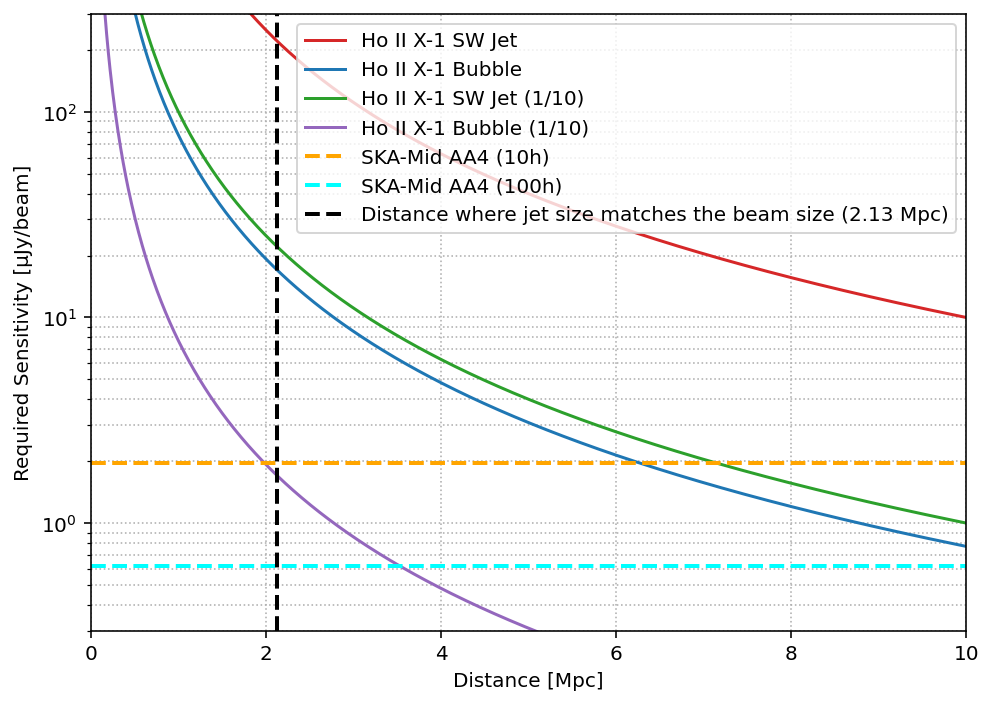}
    \caption{Required sensitivity for detecting polarized emission as a function of distance. The solid lines represent four cases, with the required sensitivity normalized to the polarized intensity of the Holmberg II X-1 SW jet and nebula at 3.39 Mpc (assuming 20\% fractional polarization and a signal-to-noise ratio of 3), as well as hypothetical sources having one-tenth of these reference polarized intensities. Dashed lines mark the continuum sensitivity achievable by SKA-Mid AA4 in 10 hours (orange, 1.96 $\mu$Jy beam$^{-1}$) and 100 hours (cyan, 0.62 $\mu$Jy beam$^{-1}$) of integration time. Calculations assume an observing bandwidth of 30 MHz and SKA-Mid Band 2 (950–1760 MHz, centered at 1310 MHz); the imaging was performed using Briggs weighting (robust=1) with a synthesized beam size of 0.87 arcsec. The required sensitivity axis is plotted on a logarithmic scale.}
    \label{fig:sensitivity_estimation}
\end{figure}

In this section, we discuss the detectability of polarized emission from ULXs assuming the SKA-Mid Array Assembly 4 (AA4) configuration. The potential polarization sources considered are the outflows—such as jets—and the surrounding bubbles (see figure \ref{fig:schematic}). As a representative case, we adopt the well-studied ULX Holmberg II X-1, from which both jet and bubble radio emissions have been detected \citep{Cseh2012, Cseh2014, Cseh2015a}.

Holmberg II X-1 is located at a distance of 3.39 Mpc. The flux density of its southwestern jet is 85 $\mu$Jy at 5.5 GHz, while the associated bubble exhibits a flux density of 613 $\mu$Jy at 4.8 GHz. The spectral indices are -1.0 for the jet and -0.53 for the bubble. The jet’s projected diameter is approximately 9 pc, and that of the bubble is 60 pc. As there have been no previous detections of ULX polarization, we assume a fractional polarization of 20\%, consistent with typical values observed in SNRs and jets. For the present sensitivity estimate, we adopt Band 2 (950–1760 MHz, centered at 1310 MHz), avoiding low frequencies strongly affected by depolarization and high frequencies with lower synchrotron emissivity. The assumed elevation is 45°, with image weighting set to Briggs and robust = 1, resulting in a synthesized beam of approximately 0.87 arcsec. Under these assumptions, the expected polarized intensities at 1310 MHz are 260 $\mu$Jy beam$^{-1}$ for the southwestern jet and 20 $\mu$Jy beam$^{-1}$ for the bubble. To achieve a signal-to-noise ratio = 3 for detection, the sensitivities required are 87 $\mu$Jy beam$^{-1}$ and 6.7 $\mu$Jy beam$^{-1}$, respectively. We consider the detectability of ULX polarization sources with these intensities over distances of 0–10 Mpc, and further examine cases that are an order of magnitude fainter than Holmberg II X-1.

Figure \ref{fig:sensitivity_estimation} plots the required sensitivity to achieve signal-to-noise ratio = 3 as a function of source distance. Red and blue lines correspond to the jet and bubble of Holmberg II X-1, while green and purple lines indicate emissions ten times fainter than those, respectively. The orange dashed line shows the achievable sensitivity for 10 hours of SKA-Mid AA4 Band 2 observations under the above assumptions, and the cyan dashed line represents 100 hours. For ULXs with jet intensities comparable to that of Holmberg II X-1, the polarized emission should be detectable within observation times shorter than $\sim$10 hours, even if the source resides in a galaxy more than 10 Mpc away. It should be noted, however, that resolving the jet structure spatially will only be feasible for nearby sources. Given the 0.87 arcsec beam, the jet lobes of Holmberg II X-1 would match the beam size already at a distance of 2.13 Mpc. Thus, resolving the internal structure requires observations of nearby ULXs or higher-resolution imaging, such as through VLBI including SKA antennas. 
In contrast, the bubble around Holmberg II X-1 is significantly larger; the beam size becomes comparable to its diameter at a distance of 14.2 Mpc. Consequently, structural resolution remains possible for sources closer than this. Although bubbles are fainter than jets and thus require longer integrations, 100 hours of observation with SKA-Mid should allow detection of bubbles associated with ULXs in galaxies up to approximately 10 Mpc away.

For sources possessing polarized flux densities an order of magnitude lower than those of Holmberg II X-1, the jet component remains detectable within 10 Mpc under realistic integration times, though spatial resolution is again limited to nearby objects. The detection of bubble polarization, however, would be feasible only for very nearby systems within roughly 3.5 Mpc, even with 100 hours of integration.

In summary, although several challenges remain even with the capabilities of SKA-AA4, the first detection of polarized emission from a ULX—and the potential to spatially resolve its structure—appears within reach. Such observations would represent a major step toward understanding the magnetic field geometry and jet–ISM interactions in these extreme accreting systems.

The detectability estimates presented above assume a fractional polarization of 20\%, which is typical for synchrotron-emitting sources such as supernova remnants and jets. However, the intrinsic polarization fraction in ULXs is currently unknown and may be significantly lower due to turbulent magnetic fields and internal Faraday depolarization. 
To assess the impact of this uncertainty, we consider a more conservative scenario with p = 5\%. Since the polarized intensity scales linearly with p, the required sensitivity increases by a factor of four compared to the 20\% case, and the required observing time increases by approximately a factor of sixteen. Under this assumption, the detection of jet polarization in Holmberg II X-1–like systems remains feasible within realistic integration times, particularly for nearby sources. In contrast, detecting polarized emission from extended bubble structures becomes significantly more challenging and may be limited to the nearest systems or to regions with enhanced magnetic field ordering, such as shock-compressed interfaces.
These results highlight that the assumed polarization fraction is a key parameter in determining the feasibility of ULX polarization studies, and future observations will be essential to constrain its typical value.

It is also instructive to compare the expected performance of SKA-Mid with that of existing radio facilities such as MeerKAT and the Karl G. Jansky Very Large Array (VLA). While both instruments have enabled significant progress in studies of faint radio sources, their sensitivity limits pose challenges for polarization measurements of ULXs at distances of several Mpc.
For typical integration times, the continuum sensitivity of MeerKAT and the VLA is of the order of a few $\mu$Jy beam$^{-1}$, whereas SKA-Mid is expected to reach sub-$\mu$Jy sensitivity under comparable observing conditions. Given that polarization detection requires higher signal-to-noise ratios and that the polarized intensity is only a fraction of the total emission, this improvement is critical.
As shown in our estimates, even for relatively bright ULX systems such as Holmberg II X-1, the expected polarized intensity of extended structures such as bubbles is at the level of a few tens of $\mu$Jy beam$^{-1}$ or lower. Detecting such signals with current facilities would require prohibitively long integration times and would be further complicated by depolarization effects and limited angular resolution.
In contrast, the enhanced sensitivity and wide bandwidth of SKA-Mid Band 2 will enable not only the first detections of polarized emission from ULXs but also the application of techniques such as Faraday tomography. Therefore, while existing facilities may marginally detect polarization in the brightest and nearest systems, SKA will be uniquely capable of conducting systematic and physically informative polarization studies of ULXs.

For completeness, we briefly compare the expected performance between earlier SKA configurations (e.g., AA*) and the AA4 configuration adopted in this study. The continuum sensitivity of an interferometric array scales approximately as $\sigma \propto \left[N_{\rm ant}(N_{\rm ant}-1)\right]^{-1/2}$, where $N_{\rm ant}$ is the number of antennas. Therefore, the increased number of antennas in the AA4 configuration leads to a sensitivity improvement of a factor of a few compared to earlier configurations such as AA*.
This improvement is particularly important for polarization studies, since the polarized intensity is only a fraction of the total emission and requires higher signal-to-noise ratios for detection. In earlier configurations such as AA*, the detection of polarized emission from ULXs would be largely limited to the brightest and most compact jet components, while extended and fainter structures such as bubbles would remain challenging to detect within practical observing times.
In contrast, the enhanced sensitivity of AA4, especially in Band 2, enables the detection of polarized emission from ULX jets in a robust manner and makes the detection of bubble components feasible in favorable cases. This represents a transition from feasibility-limited observations to a regime where systematic and physically informative polarization studies of ULXs become possible.

\section{Conclusion}
Ultraluminous X-ray sources (ULXs) are among the brightest non-nuclear X-ray emitters in nearby galaxies and represent key laboratories for studying extreme accretion physics. Their origin has been attributed primarily to two scenarios: sub-Eddington accretion onto intermediate-mass black holes (IMBHs) and supercritical accretion onto stellar-mass compact objects in binary systems. In either case, ULXs release enormous energy through both radiative and mechanical processes, often driving powerful radio-emitting jets and bubble-like nebulae in their surrounding medium. While radio emission has been detected from approximately fifteen ULXs, polarization has not yet been observed, largely due to limitations in observational sensitivity and angular resolution at several-megaparsec distances.

Using the performance parameters of SKA-Mid in its AA4 configuration, we estimated the detectability of polarized emission from a ULX jet and bubble under realistic assumptions. Our analysis shows that a Holmberg II X-1–like jet can be detected in polarized intensity within about 10 hours of integration for systems at distances up to approximately 10 Mpc, while bubble polarization detection is feasible with 100 hours of observation. However, spatially resolving detailed structures will remain challenging except for nearby sources, emphasizing the need for higher angular resolution through future SKA-based VLBI observations.

These results suggest that the SKA will enable the first polarimetric detections of ULXs, marking a major step forward in our understanding of their magnetic field structure and jet–ISM interactions. Polarimetric studies with the SKA will not only clarify the magnetic geometry and turbulence conditions associated with ULX outflows and bubbles but also provide critical insights into the physics of super-Eddington accretion, jet launching, and cosmic-ray acceleration in extragalactic environments.











\bibliographystyle{abbrvnat-maxbibnames4}
\bibliography{chapter} 

\end{document}